# Iron Isotope Effect in SmFeAsO$_{0.65}$ and SmFeAsO$_{0.77}$H$_{0.12}$ Superconductors: A Raman Study

Birender Singh[1], P. M. Shirage[2], A. Iyo[3] and Pradeep Kumar[1*]

[1]School of Basic Sciences, Indian Institute of Technology Mandi, Mandi-175005, India

[2]Metallurgy Engineering and Materials Science & Physics, Indian Institute of Technology Indore, Indore-453552, India

[3]National Institute of Advanced Industrial Science and Technology, Tsukuba, Ibaraki 305-8568, Japan.

## ABSTRACT

We report the inelastic light scattering studies on SmFeAsO$_{0.65}$ and SmFeAsO$_{0.77}$H$_{0.12}$ with iron isotopes namely $^{54}$Fe and $^{57}$Fe. In both of these systems under investigation we observed a significant shift in the frequency of the phonon modes associated with the displacement of Fe atoms around ~ 200 cm$^{-1}$. The observed shift in the Fe mode (B$_{1g}$) for SmFeAsO$_{0.65}$ is ~ 1.4 % and lower in case of SmFeAsO$_{0.77}$H$_{0.12}$, which is ~ 0.65 %, attributed to the lower percentage of isotopic substitution in case of SmFeAsO$_{0.77}$H$_{0.12}$. Our study reveals the significant iron isotope effect in these systems hinting towards the crucial role of electron-phonon coupling in the pairing mechanism of iron based superconductors.

[*] Corresponding author: email id: khatri0003@gmail.com

# 1. INTRODUCTION

The discovery of high temperature superconductivity in iron based superconductors LaFeAs$_{1-x}$F$_x$ with $T_c$ = 26 K [1], has provided challenging as well as exciting new avenues in the field of superconductivity to explore these materials both experimentally and theoretically. This class of superconducting materials has attracted a lot of attention because of the quite high superconducting transition temperature, despite Fe being one of its constituents. Immediately after this discovery, higher temperature superconductors were discovered by replacing La with other rare earth elements, i.e. Sm [2-3], Ce [4], Nd [5-7], Pr [8] and Gd [9] in LaFeAsO$_{1-x}$F$_x$ system with $T_c$ ~ 56 K. Since then continuous efforts have been made to understand these high temperature iron based superconducting materials. Consequently, several families of iron-based superconductors (FeBS) have been discovered, such as 1111-type i.e. LnFeAsO$_{1-x}$F$_x$ (Ln = lanthanide elements) [10-13], 122-type i.e. AFe$_2$As$_2$ (A = Ba, Ca, Sr) [14-18], 111-type i.e. LiFeAs [19-21], LiFeP [22] and NaFeAs [23], 11-type i.e. FeTe$_{1-x}$Se$_x$ [24-27] and Ba$_4$Sc$_3$O$_{7.5}$Fe$_2$As$_2$ and Ca$_4$Al$_2$O$_{6-x}$Fe$_2$As$_2$ [28-29]. Very recently, a lot of effort has been made to investigate the thin films of FeBS. Interestingly, the transition temperature in thin films is reported to be as high as 100 K, significantly higher than the bulk systems [30] crossing the liquid nitrogen bar of 77 K.

Although, the superconducting transition temperature in these systems has been elevated as high as 100 K, the mechanism for superconductivity still remains an open question. The parent compounds exhibit long-range anti-ferromagnetic ordering, which is suppressed on doping and superconductivity emerges. Recent experimental as well as theoretical studies have suggested the strong interplay between different degrees of freedom such as spin and orbital which may be responsible for the elusive pairing mechanism in these systems [31-35]. The earlier estimated value of the electron-phonon coupling within the Eliashberg theory was found to be very weak to generate the required $T_c$ for these systems [36-38], and as a result an unconventional origin of superconductivity mediated by AFM spin-fluctuations was proposed [39]. In another study, the effect of magnetisation and doping was calculated on the electron-phonon coupling, and it was found that electron-phonon coupling strength increases significantly via spin-channels [40-42]. Theoretical calculations have also suggested that electronic structure of FeBS systems is strongly perturbed below the magnetic transition temperature due to the opening of spin density wave gap [43-45]. The electron-phonon coupling increases substantially for states at the Fermi surface due to anisotropic coupling between phonons and magnetic moments (i.e. magneto-elastic coupling) in the AFM structure below structural and magnetic transition temperature. The strength of the electron-phonon coupling constant, calculated by taking into account the many body effects is found to be $\lambda_{e\text{-}p}$ ~1, which is much larger than that calculated using LDA band structure calculation ($\lambda_{e-p}^{LDA}$ ~ 0.2), suggesting that the role of electron-phonon coupling can not be neglected in the pairing mechanism. In fact, the electron-phonon strength in FeBS is compared to that in $MgBr_2$ superconductor, where the superconductivity is mediated by electron-phonon coupling. In addition to these theoretical calculations, a large isotope coefficient as high as $\alpha_{Fe}$ ~ 0.81 has been reported experimentally for different families of FeBS [46-52], clearly suggesting that the electron-phonon coupling can not be ruled out in understanding the pairing mechanism in these materials along with other degrees of freedom such as spin and orbital.

Existing reports in the literature, with finite iron isotope coefficient in different families of FeBS, suggest that the study of isotope effect is pertinent to investigate the role of electron-phonon coupling in these materials. In the present work, we have undertaken such a study and have investigated the iron isotope effect by exchanging $^{54}$Fe and $^{57}$Fe on oxygen deficient $SmFeAsO_{0.65}$ and hydrogen-doped $SmFeAsO_{0.77}H_{0.12}$ systems via inelastic Raman scattering measurements. A large frequency shift of phonon modes associated with displacement of Fe ($B_{1g}$) atoms is observed for both SmFeAsO$_{0.65}$ and SmFeAsO$_{0.77}$H$_{0.12}$ with the exchange of isotopes; however, other modes are not affected by the isotope substitution, suggesting that the percentage of constituents is same in different isotope substituted systems. We also observed that the frequency shift after exchanging isotopes is more in case of SmFeAsO$_{0.65}$ as compared to SmFeAsO$_{0.77}$H$_{0.12}$. Our experimental results evidenced the strong isotope effect in these systems suggesting the crucial role of electron-phonon coupling in understanding the underlying pairing mechanism responsible for the resistance free flow of the charge carriers in these materials.

## 2. EXPERIMENTAL DETAILS

Polycrystalline samples of SmFeAsO$_{0.65}$ ($T_c$ ~ 52 K) and SmFeAsO$_{0.77}$H$_{0.12}$ ($T_c$ ~ 54 K) with nominal composition were synthesized and characterized as described in reference [48]. The percentage of iron isotope ($^{57}$Fe /$^{54}$Fe) in case of SmFeAsO$_{0.77}$H$_{0.12}$ is 43 %, whereas in case of SmFeAsO$_{0.65,}$ it is 100%. Unpolarized micro Raman measurements were carried out in backscattering geometry using He-Ne ($\lambda$ = 633 nm) and Raman spectrometer (LabRam HR-Evolution) coupled with a Peltier cooled CCD and spectrometer with 1800 lines/mm grating. Laser power at the sample was kept very low (~ 1mW) to avoid any heating of the sample.

## 3. RESULTS AND DISCUSSIONS

SmFeAsO has a layered structure belonging to the tetragonal P4/nmm space group, containing two SmFeAsO units per unit cell. Out of 24 normal modes at the $\Gamma$ point eight are Raman active phonon modes belonging to the irreducible representation 2A$_{1g}$ + 2B$_{1g}$ + 4E$_g$. Figure 1 (a and b) and Figure 2 (a and b) show the room temperature Raman spectrum of SmFeAsO$_{0.65}$ ($^{54}$Fe and $^{57}$Fe) and SmFeAsO$_{0.77}$H$_{0.12}$ ($^{54}$Fe and $^{57}$Fe), respectively, revealing four modes labeled as S1 to S4 in the spectral range of 140 - 270 cm$^{-1}$. Spectra are fitted with sum of Lorentzians to extract the peak frequencies. The individual modes are shown by thin lines and resultant fit by a thick line. The average frequencies, as we have recorded Raman spectra at many points for each sample and have taken the average, of phonon modes for four different samples are tabulated in Table-I. Following the earlier Raman studies on "1111" systems [53-60], and based on our observations, we assign the observed phonon modes as S1: ~ 165 cm$^{-1}$ ( $A_{1g}$, Sm ); S2 ~ 200 cm$^{-1}$ ( $B_{1g}$, Fe ); S3 ~ 225 cm$^{-1}$ ( $A_{1g}$, As ); and S4 ~ 250 cm$^{-1}$ ( $E_g$, Fe and As ).

Figure 3 shows the comparative Raman spectra of SmFeAsO$_{0.65}$ with $^{54}$Fe and $^{57}$Fe substitution. We observed a clear shift in the phonon mode associated with the Fe - $B_{1g}$ ( ~ 200 cm$^{-1}$ ) mode. The observed shift ($\Delta\omega$) in the phonon frequency is ~ 2.8 cm$^{-1}$ ( ~ 1.4 % ). This large shift in the mode frequency suggests a significant isotope effect in these class of superconductors. We note that a similar shift in the mode associated with Fe atoms displacement was also observed in the fluorine doped FeBS [46]. We also performed similar studies on H-doped system, i.e. SmFeAsO$_{0.77}$H$_{0.12}$. Superconducting transition temperature in these systems depends strongly on the *a*-lattice parameter, and maximum $T_c$ is achieved at optimised *a*-lattice parameter [48]. Optimised *a*-lattice leads to a significant increase in the transition temperature by ~ 15 K as compared to the under doped samples. The effect of hydroxide doping that also leads to optimised *a*-lattice parameter, motivated us to study the Fe isotope effect in this system as well. Figure 4 shows the comparative spectra of SmFeAsO$_{0.77}$H$_{0.12}$ with $^{54}$Fe and $^{57}$Fe substitution. We note that percentage of $^{54}$Fe and $^{57}$Fe is only 43 %. We observed a clear shift in the Fe - $B_{1g}$ mode i.e. $\Delta\omega$ ~ 1.3 cm$^{-1}$ ( ~ 0.65 % ) (see Table-II). Our observation of clear shift in the Fe ($B_{1g}$) mode frequency in both of the systems suggests the crucial role of iron isotope in these superconducting materials.

The isotope coefficient in these systems may be defined similar to the $T_c$ isotope coefficient ($-\frac{M}{\Delta M}\frac{\Delta T_c}{T_c}$) using the Einstein temperature, where the Einstein temperature ($T_E$) is defined as $T_E = \sqrt{\frac{k_0}{\mu}}$, $k_0$ is the effective force constant and $\mu$ is the reduced mass [51]. We introduced the isotope coefficient ($\alpha^\omega$) using the phonon frequency as the phonon frequency is intimately related to the Einstein temperature via effective force constant $k_0 \propto \omega^2$ so $T_E \propto \omega$. Therefore, $\alpha^\omega$ may be defined as $-\frac{M}{\Delta M}\frac{\Delta \omega}{\omega}$, where $\Delta \omega$ is the change in phonon frequency upon isotope substitution and M is the atomic mass of the isotope substituted. Using the above definition of isotope coefficient introduced via phonon frequency, one may get an independent evaluation of the iron-isotope effect. Using the change in Fe mode phonon frequency upon isotope substitution given in Table-I, we estimated the value of $\alpha^\omega$ as ~ 0.27 and ~ 0.28 for SmFeAsO$_{0.65}$ and SmFeAsO$_{0.77}$H$_{0.12}$, respectively (see Table-II). We note that the estimated value is much higher than the iron isotope coefficient calculated using $T_c$ on the SmFeAsO$_{0.77}$H$_{0.12}$ system [48]. Khasanov et al.,[50] pointed out that isotope substitution in these systems causes a small structural modification and that inturn affects $T_c$. They did a detailed analysis of the existing studies on the iron-isotope effect using $T_c$ and found that the intrinsic isotope effect, which is independent of structural changes, in all of these systems is consistent and of the order of ~ 0.35 - 0.4. Our estimated value of the intrinsic isotope effect (using phonon frequency) for both of these systems under investigation is also of the same magnitude. Therefore, the observed change in the phonon frequency in both of these superconducting systems upon isotope substitution and estimated isotope coefficient clearly suggest the intrinsic nature of iron-isotope coefficient in these systems. In a theoretical report by Bussmann-Holder et al., [61] on different FeBS systems based on the electron-phonon mediated superconductivity within the dominant gap channel predicted $T_c$ independent isotope effect and estimated isotope coefficient $\alpha$ of the order of 0.5. We also note that FeBS undergoes symmetry breaking from tetragonal to orthorhombic phase associated with the orbital ordering which is accompanied by magnetic phase transition, highlighting the underlying role of coupled electronic and magnetic degrees of freedom. Our observation of Fe isotope effect, in line with earlier studies on different families of FeBS, clearly point towards the intricate role of phononic degrees of freedom coupled with electronic and magnetic degrees of freedom.

## 4. CONCLUSIONS

In the present work, we have performed Raman measurements on $SmFeAsO_{0.65}$ and $SmFeAsO_{0.77}H_{0.12}$ superconductors at room temperature. The significant change in the phonon frequency of the Fe - $B_{1g}$ mode ( ~ 200 cm$^{-1}$ ) is observed by replacing iron isotope $^{54}$Fe with $^{57}$Fe in both $SmFeAsO_{0.65}$ and $SmFeAsO_{0.77}H_{0.12}$, suggesting the strong iron isotope dependence of the phonon modes associated with the displacement of Fe atoms. The large change in the phonon frequency upon isotope substitution suggests the intrinsic isotope effect in these systems. Our observations suggests that the lattice vibrations do play an important role in providing the glue for the Cooper pair formation and their role can not be ignored. Therefore, the phononic degrees of freedom should be considered at par with other degrees of freedom, such as electronic and magnetic, to unravel the underlying mechanism for the superconductivity.

## ACKNOWLEDGMENTS

PK acknowledgs the Department of Science and Technology, India, for INSPIRE fellowship and Advanced Material Research Center, IIT Mandi, for the experimental facilities. PMS acknowledges SERB-DST for awarding the Ramanujan Fellowship (SR/S2/RJN-121/2012).

**Table-I:** Room temperature average peak frequencies ( in cm$^{-1}$ ) for the Raman bands of SmFeAsO$_{0.65}$ and SmFeAsO$_{0.77}$H$_{0.12}$ with $^{54}$Fe and $^{57}$Fe isotope substitutions.

| Mode Assignment | Frequency (cm$^{-1}$) (SmFeAsO$_{0.65}$) | | Frequency (cm$^{-1}$) (SmFeAsO$_{0.77}$H$_{0.12}$) | |
|---|---|---|---|---|
| | $^{54}$Fe | $^{57}$Fe | $^{54}$Fe | $^{57}$Fe |
| S1 (A$_{1g}$ - Sm) | 165 | 164.9 | 164.4 | 164.8 |
| S2 (B$_{1g}$ - Fe) | 198.4 | 195.6 | 198.2 | 196.9 |
| S3 (A$_{1g}$ - As) | 224.3 | 224.8 | 225.8 | 225.7 |
| S4 (E$_g$ - Fe and As) | 250.8 | 251.2 | 251.4 | 252 |

**Table-II:** Raman shift ($\Delta\omega$), in cm$^{-1}$, for the Fe - B$_{1g}$ mode and estimated isotope effect using phonon frequency ($\alpha^{\omega}$). For the calculation, we have used the molar masses for isotope exchange as 56.94 g and 53.94 g for $^{57}$Fe and $^{54}$Fe, respectively, for SmFeAsO$_{0.65}$; and 56.32 and 55.02 for for $^{57}$Fe and $^{54}$Fe, respectively, for SmFeAsO$_{0.77}$H$_{0.12}$ [48].

| | Raman Shift ($\Delta\omega$), in cm$^{-1}$, for the Fe - B$_{1g}$ mode | $\alpha^{\omega}$ |
|---|---|---|
| SmFeAsO$_{0.65}$ | 2.8 | 0.27 |
| SmFeAsO$_{0.77}$H$_{0.12}$ | 1.3 | 0.28 |

**FIGURE CAPTION:**

**Figure 1:** (Color online) Room temperature Raman spectra for SmFeAsO$_{0.65}$ with $^{54}$Fe (a) and $^{57}$Fe isotope (b), showing four Raman active phonon modes labelled as S1 to S4. The mode are assigned as S1: ~ 165 cm$^{-1}$ (A$_{1g}$, Sm); S2 ~ 200 cm$^{-1}$ (B$_{1g}$, Fe); S3 ~ 225 cm$^{-1}$(A$_{1g}$, As); and S4 ~ 250 cm$^{-1}$(E$_g$, Fe and As).

**Figure 2:** (Color online) Room temperature Raman spectra for SmFeAsO$_{0.77}$H$_{0.12}$ with $^{54}$Fe (a) and $^{57}$Fe isotope (b), showing four Raman active phonon modes labelled as S1 to S4.

**Figure 3:** (Color online) Comparison of Raman spectra for SmFeAsO$_{0.65}$ with iron isotope $^{54}$Fe (filled sphere) and $^{57}$Fe (circle). The vertical dashed lines are guide to the eyes showing the S2 (Fe-B$_{1g}$) mode shift upon isotope substitution.

**Figure 4:** (Color online) Comparison of Raman spectra of SmFeAsO$_{0.77}$H$_{0.12}$ for iron isotope $^{54}$Fe (filled sphere) and $^{57}$Fe (circle). The vertical dashed lines are guide to the eyes showing the S2 (Fe-B$_{1g}$) mode shift.

**Figure 1:**

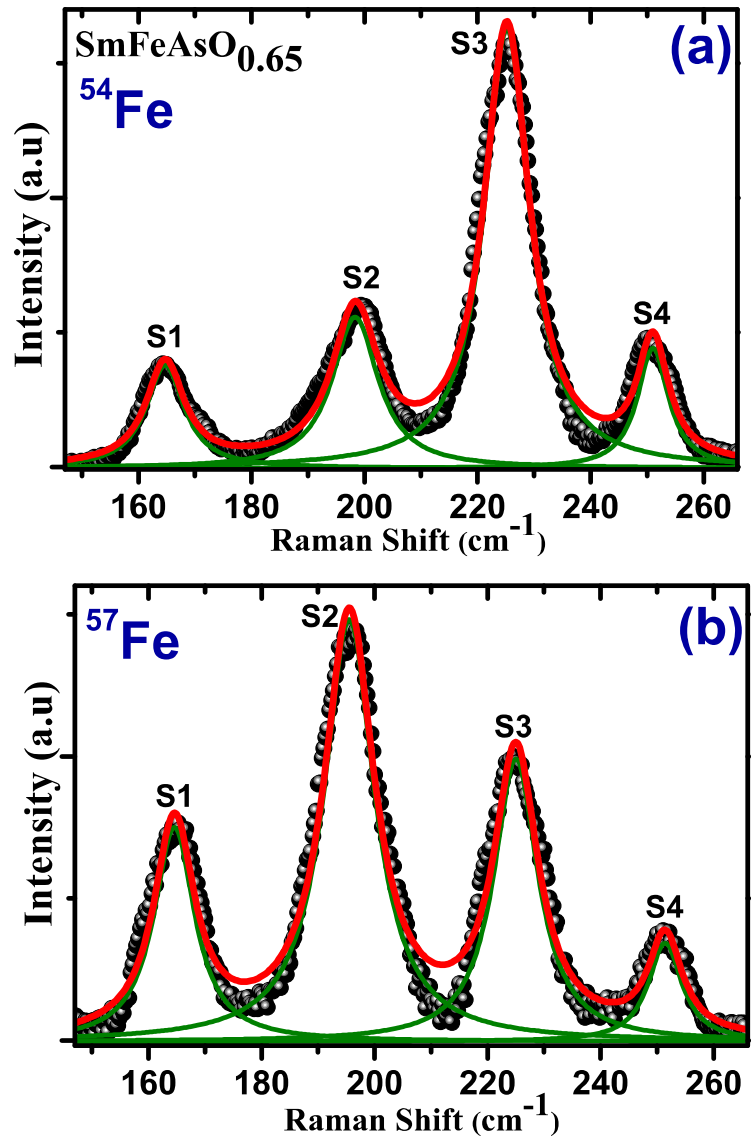

Figure 2:

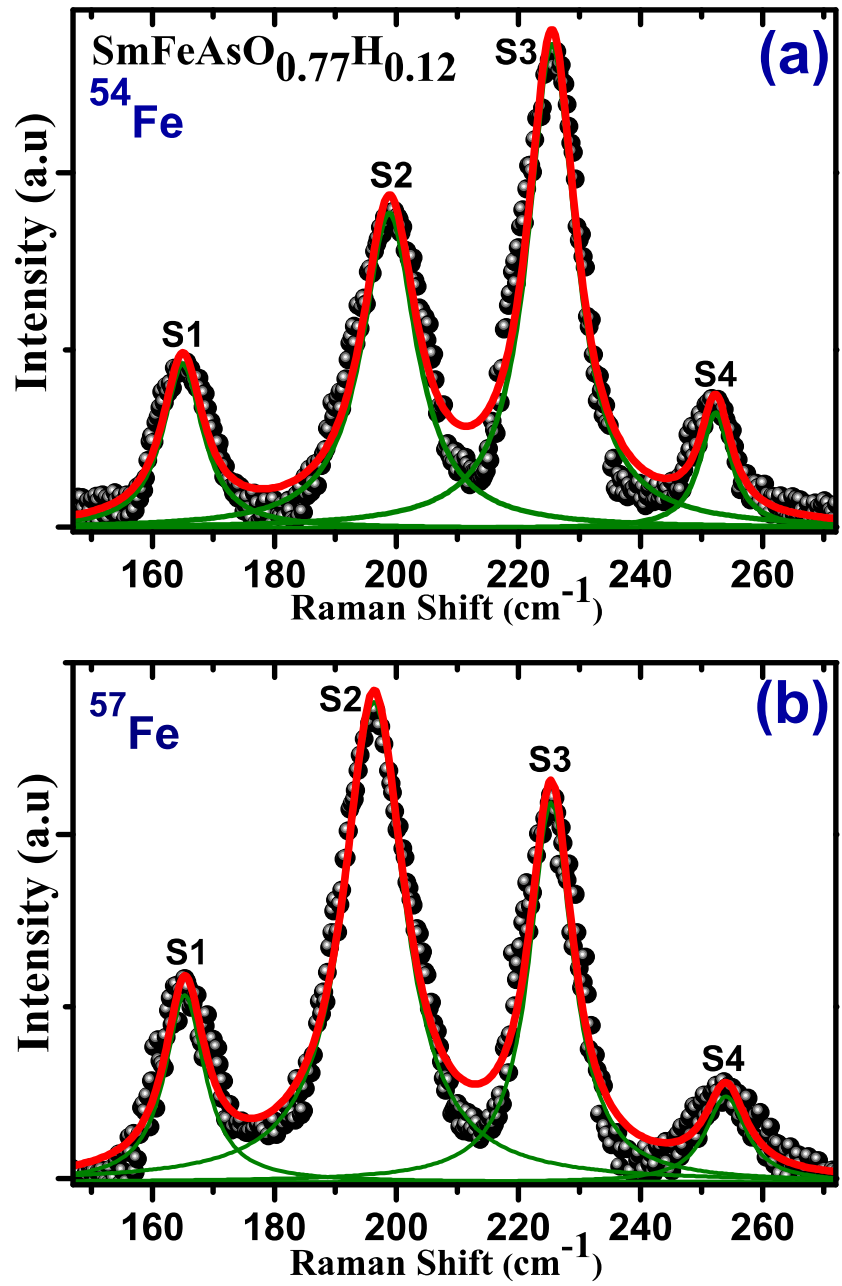

**Figure 3:**

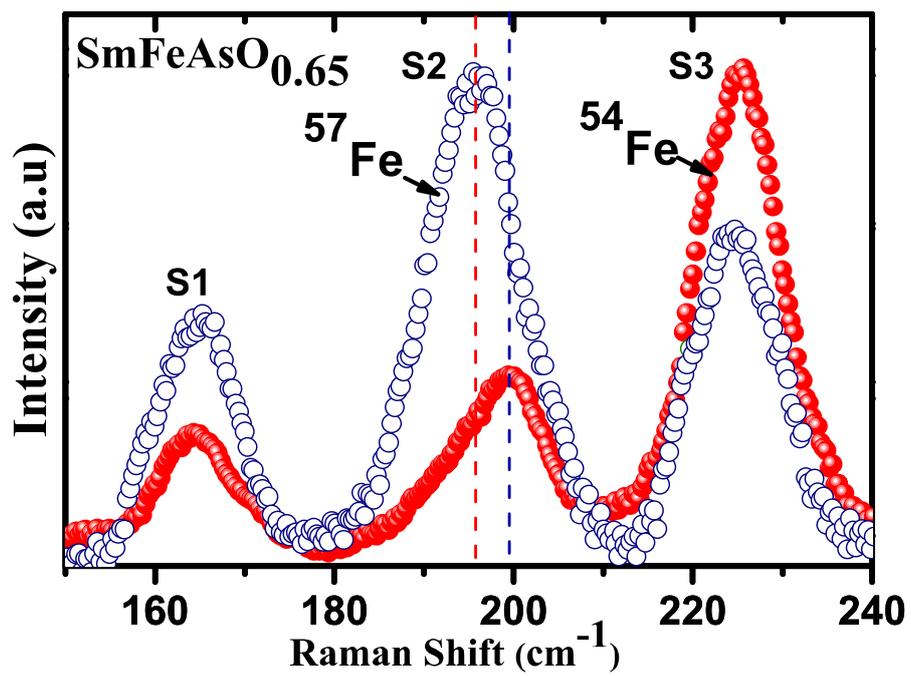

**Figure 4:**

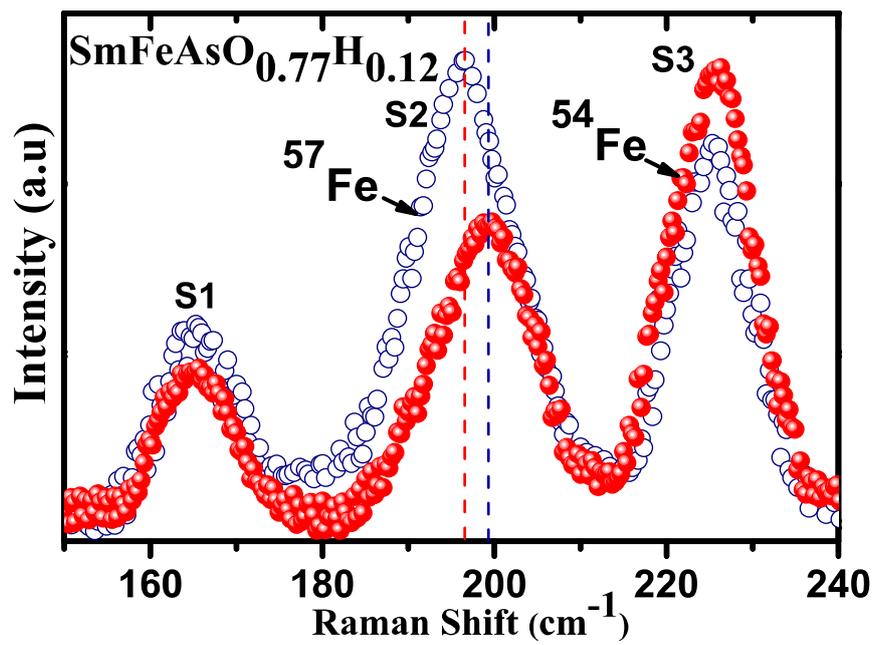